\documentclass[prb,twocolumn,amssymb,showpacs]{revtex4}
\usepackage{graphicx}
\usepackage{amsmath}

\begin{document}

\title{ESR of MnO embedded in silica nanoporous matrices with different topologies.}
\author{I. V. Golosovsky}
\affiliation{Petersburg Nuclear Physics Institute, 188300, Gatchina, St. Petersburg, Russia.}
\author{D. Ar\v{c}on, Z. Jagli\v{c}i\v{c}, and P. Cevc}
\affiliation{Institute "Jozef Stefan", 1000, Ljubljana, Slovenia.}
\author{V. P. Sakhnenko}
\affiliation{Rostov State University, 344090, Rostov/Don, Russia.}
\author{D. A. Kurdyukov and Y. A. Kumzerov}
\affiliation{A. F. Ioffe Physico-Technical Institute, 194021, St. Petersburg, Russia.}

\begin{abstract}
Electron spin resonance (ESR) experiments were performed with antiferromagnetic MnO confined within a porous
vycor-type glass and within MCM-type channel matrices. A signal from confined MnO shows two components from
crystallized and amorphous MnO and depends on the pore topology. Crystallized MnO within a porous glass shows a
behavior having many similarities to the bulk. In contrast with the bulk the strong ESR signal due to
disordered "surface" spins is observed below the magnetic transition. With the decrease of channel diameter the
fraction of amorphous MnO increases while the amount of crystallized MnO decreases. The mutual influence of
amorphous and crystalline MnO is observed in the matrices with a larger channel diameter. In the matrices with
a smaller channel diameter the ESR signal mainly originates from amorphous MnO and its behavior is typical for
the highly disordered magnetic system.

\end{abstract}

 \maketitle

\section{Introduction}

The magnetic properties of nanoparticles are of great interest because of their unique physical properties and
practical applications. In particular, the magnetic properties drastically change when the particle size
becomes comparable with the length of the magnetic interaction or the length of spin diffusion.

The present work addresses these effects by studying the classical antiferromagnet MnO, confined to porous
silica matrices of different topologies and sizes of voids. We used a porous glass \cite{Levitz} with a random
network of pores and novel mesoporous matrices \cite{Grun,Zhao} with a system of regular nanochannels. The
morphology of the MnO particles thus varies considerably depending on the matrix and can emphasize different
dimensionalities of the spin system.

We have reported  that confined MnO reveals some remarkable differences when compared to the bulk
\cite{MnO-PRL,neutrons}. Below the magnetic transition a "long range magnetic ordering" appears within a "core"
with strongly reduced magnetic moment. A first order magnetic transition at 118 K in the bulk becomes a second
order transition in confinement with the N\'eel temperatures ($T_N$) of 120-128 K  depending on the host
matrix.

Diffraction experiments \cite{LURE} show that confined MnO exists within the channels in crystallized and in
amorphous forms. The amorphous MnO manifests itself as specific diffuse scattering, which overlaps with the
diffuse background scattering due to an amorphous silica matrix and therefore they cannot be separated. The
magnetic properties of a amorphous fraction of MnO are thus completely unexplored and other experimental
techniques are needed.

Obviously, the effect of pore size and shape of the nanoparticles should be investigated in a more systematic
way. The goal of the present work is to shed light on the differences in the magnetic properties of MnO
confined within nanopores with different topologies.

In the present study an electron spin resonance (ESR) technique is used to probe magnetic properties of MnO
confined within channel type matrices and within a porous glass. The ESR technique is often used for
characterizations of the different paramagnetic centers or complexes embedded into the nanochannel materials
\cite{Xu} as well as nanomagnets \cite{Barra} and it is very powerful method for investigations of spin
dynamics in paramagnetic phases. In the present work we give an interpretation of the ESR data on the basis of
the recent synchrotron diffraction studies of the same samples \cite{LURE}.

The ESR technique enabled us to study the development of magnetic correlations in "crystallized" as well as in
amorphous fractions of confined MnO. In particular we found remarkable peculiarities in confined MnO, which
demonstrate the importance of topology and surface effects for magnetism in confinement.

\section{Samples and experiment}

The ESR experiments were performed with MnO confined to different channel type matrices known as MCM-41
\cite{Grun} or SBA-15 \cite {Zhao} and confined to a porous glass known as vycor \cite{Levitz}. For comparison
the measurements were carried out with the commercially available powder bulk MnO of high purity.

MCM-41 matrices with 24 and 35 {\AA} channel diameters and SBA-15 matrices with 47, 65 and 87 {\AA} channel
diameters were used. The MCM-41 and SBA-15 matrices (referred below as SBA and MCM) differ by the preparation
technique and both are amorphous silica (SiO$_{2}$) matrices with a regular hexagonal array of parallel
nanochannels with a wall thickness of $\sim$ 8-10 {\AA}. The matrices in the form of a powder with a grain size
of $\sim$ 1-2 ${\mu}m$ were prepared in the Laboratoire de Chimie Physique, Universit\'e Paris-Sud, France
\cite{Morineau-2}.

All samples were filled with MnO from the liquid components by the "bath deposition method" with the following
crystallization inside the voids. X-ray and neutron diffraction measurements show the presence of crystallized
stoichiometric MnO\cite{LURE,neutrons}.

MnO confined within the large (47-87 {\AA}) channels crystallizes in the form of thin (about 10 {\AA}),
ribbon-like structures with a width of about a channel diameter and a length of 180-260 {\AA}. In the matrices
with the narrower channels (24 and 35 {\AA}) MnO crystallizes in the form of nanowires with diameters of $\sim
$ 20 {\AA} and a length of $\sim $ 180-200 {\AA} \cite{LURE}.

Unlike channel type matrices a vycor glass has a random interconnected network of elongated pores with a narrow
(only few percent) distribution of pore diameters about 70 {\AA} \cite{Levitz}. It has been demonstrated that
MnO confined to a porous glass forms aggregates with the average diameter of 145(3) {\AA} \cite{MnO-PRL}.

Samples in a powder form, (apart of a porous glass), around 5 $mg$, were transferred to the quartz tubes in Ar
atmosphere. Before measurements the samples were evacuated for several hours at a temperature of 100
C$^{\circ}$ to avoid moisture contamination after which they were sealed.

ESR experiments were performed with a Bruker E580 spectrometer operating in X-band (9.63 GHz) at the Institute
"Jozef Stefan", Ljubljana, Slovenia. Small changes in the Q-factor of the cavity (which might slightly affect
the measured susceptibility of the sample) were corrected with a reference sample in a second cavity. The
Oxford Instruments continuous flow cryostat ESR900 combined with a temperature controller ITC 503 was used. The
results presented below were obtained on cooling from 250 to 20 K.

With the same samples the magnetization at applied field up to 6 T were measured using the standard SQUID
magnetometer.

\section{Results and discussion}

\subsection{The ESR signal lineshape analysis}

\begin{figure} [t]
\includegraphics* [width=\columnwidth] {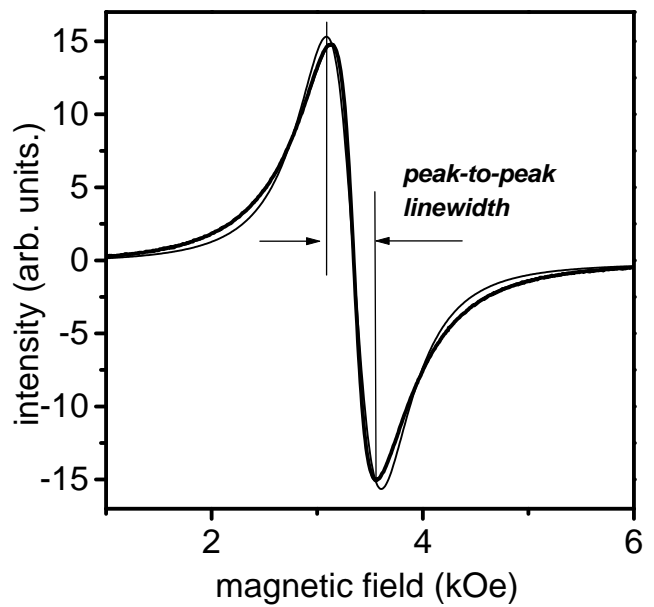}
\caption{A typical ESR signal from MnO confined within a porous glass and its fitting by one (fine line) and by
two Lorentzians (thick line coincides with the experimental curve). The ESR spectrum is displayed in a
derivative mode.} \label{signal}
\end{figure}

\begin{figure} [t]
\includegraphics* [width=\columnwidth] {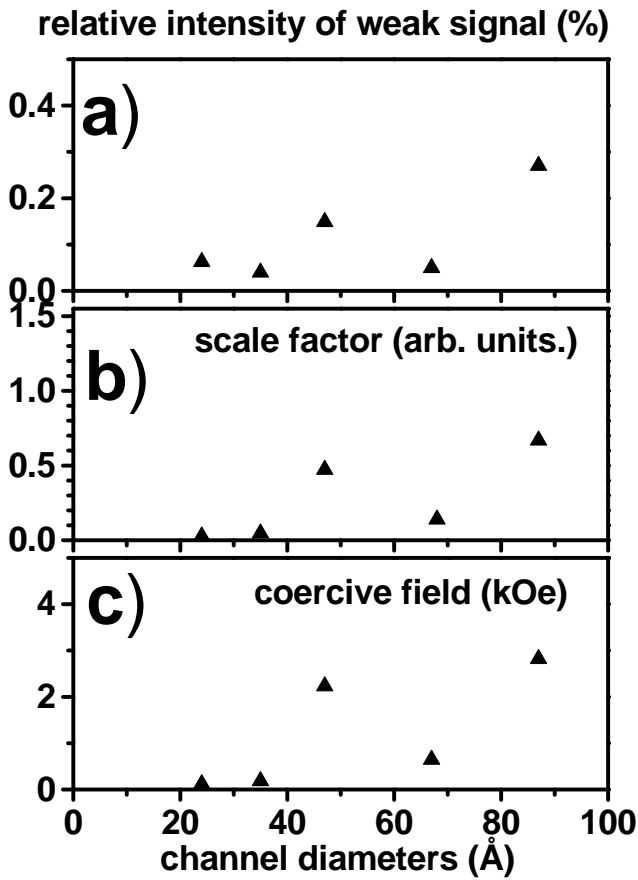}
\caption{a) Relative intensity of the second (weak) ESR signal for the channel type matrices. b) Scale factor
refined from diffraction experiments. c) The coercive field for the same samples, measured at 2 K. Errors
(e.s.d. - estimated standard deviation) do not exceed the symbol size.} \label{low-signal}
\end{figure}

In all samples we found in paramagnetic phase a very strong ESR signal (figure \ref{signal}) with a g-factor of
1.997(1), which is characteristic of $Mn^{2+}$ centers \cite{Pilbrow}. In the paramagnetic phase the g-factor
showed only a weak temperature dependence. A short-range order effects were observed only just above the N\'eel
temperature, where the g-factor value typically slightly decreased.

No noticeable signal was detected from an empty matrix. Therefore we considered the measured signal as a result
of absorption by the confined substances only.

The ESR signal from the bulk powder MnO is well approximated by a Lorentzian function. Gaussian line shape
gives much worse results. On the other hand, the fit of the ESR signal from nanostructured MnO can be
significantly improved by assuming two overlapping Lorentzians with different linewidths and intensities
(figure \ref{signal}). In this case a goodness-of-fit, a factor defining the quality of fit, diminishes by
factor of 100.

In the samples with confined MnO the independent refinement of the g-factors of two overlapping signals in the
paramagnetic region show them nearly identical that points that both signals come from $Mn^{2+}$ ions. Below
the magnetic transition when a signal decreases the refinement with the independent g-factors becomes unstable.
In the final refinement we constrained both g-factors to be the same.

There could be several different explanations for the deviation of the ESR signal from a simple single
Lorentzian function. Firstly, it is reasonable to expect that a weak ESR signal originates from confined MnO in
different states: crystallized and amorphous. The ESR signal in such a case will be a sum of two Lorentzians.

Secondly, because of a large linewidth one expects the deviations of the ESR lineshape from a Lorentzian due to
dispersion effects because of non-diagonal elements of the magnetic susceptibility \cite{Benner}. The large
linewidth are observed as for the bulk as well for confined MnO, however the signal from the bulk is well
described by one Lorentzian. Therefore we are tempted to rule out the second possibility and thus suppose that
two overlapping components are assigned to crystalline and amorphous confined MnO. This conclusion is
consistent with our diffraction studies\cite{LURE} where the diffuse scattering from amorphous MnO was
observed.

The relative intensity of the weaker signal measured for channel-type matrices at room temperature varies from
sample to sample. It is displayed in figure \ref{low-signal}a as a function of channel diameter. The
results on vycor glass are not presented in this figure because of different void topology. They will be
discussed separately.

In figure \ref{low-signal}b a scale factor vs channel diameter from  the X-ray diffraction experiments
\cite{LURE} is shown. This factor is proportional to the intensity of diffraction lines and is proportional to
the quantity of confined MnO in the crystallized state. In figure \ref{low-signal}c coercive fields for the
same samples calculated from the magnetization loops at 2 K are presented. This parameter is also proportional
to the quantity of crystallized MnO.

It is clearly seen that the dependencies of the relative intensity of a weak signal, scale factor and coercive
field are similar. Moreover, in all channel matrices the linewidth of the weak component at high temperatures
has typically a value about 1 kOe, i.e. nearly the same as in the bulk. It can be explained only by supposing
that a weak compoment results from the crystallized MnO.

These observations provide a strong evidence of the following assignment of the signal components: in the
channel type matrices the minor signal originates from confined crystallized MnO, while the main signal
originates from an amorphous MnO.

In the case of MnO confined to a porous glass (fractal voids) the signal with linewidth 1 kOe, near to that in
the bulk, dominates and shows the behavior characteristic for the bulk above the transition temperature. This
suggests that in a porous glass the main signal is due to crystalline MnO. This situation is just the opposite
to the case of channel type matrices.

It should be noted, that there is no direct correlation between described above parameters and channel
diameter, because the filling by MnO depends not only on the channel diameter but to a large degree on the
specific properties of the inner surface of the channels, in particularly on the wetting ability.

\subsection{MnO confined within a porous glass matrix, comparison with the bulk}

\begin{figure} [t]
\includegraphics* [width=\columnwidth] {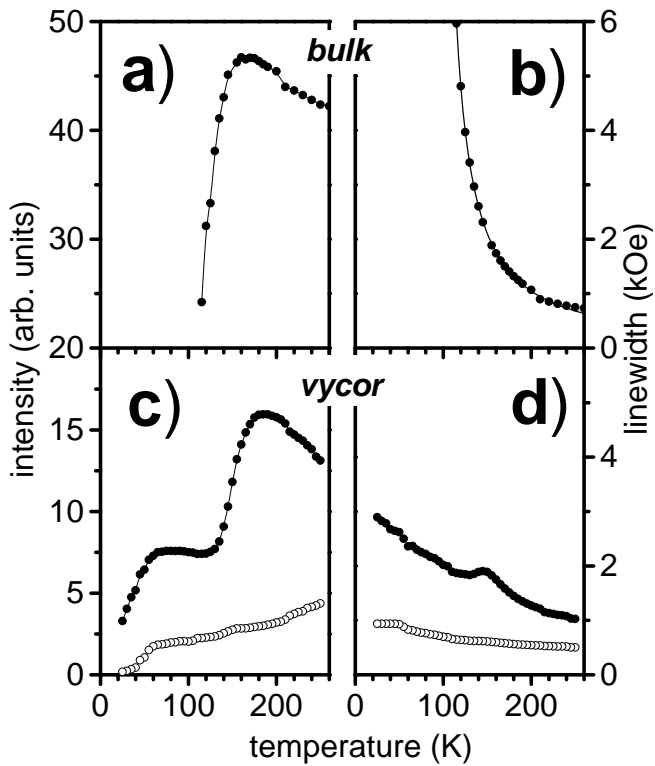}
\caption{Intensity (a) and linewidth (b) of the ESR signal from the bulk MnO. Intensity (c) and linewidth (d)
of the ESR signal from MnO confined to a porous glass. Open circles correspond to the signal from crystalline
MnO and solid circles correspond to the signal from amorphous MnO. Errors (e.s.d.) do not exceed the symbol
size.} \label{bulkvycor}
\end{figure}

In figures \ref{bulkvycor}a and \ref{bulkvycor}b the intensity of the ESR signal, which is proportional to the
spin susceptibility, and its linewidth, and which is given by the relaxation rate, are shown for the bulk MnO
versus temperature. A typical behavior for an antiferromagnet, which undergoes the magnetic transition at 118 K
is observed. On approaching $T_N$ from above, the ESR signal dramatically broadens, the linewidth diverges and
the signal disappears. In the bulk MnO a magnetic correlation length is essentially smaller than a grain size
of about some microns and there are no restrictions for the development of spin correlations and magnetic order
below $T_N$. This is in striking contrast with confined MnO, where a correlation length is constrained by the
nanoparticle size.

MnO synthesized within porous glass has a form of isotropic nanoparticles. The ESR signal was found to deviate
from a simple Lorentzian form and was thus analyzed with two overlapping Lorentzians as described above. In
figures \ref{bulkvycor}c,d the intensities and linewidths vs temperature of both components are shown. The weak
signal component with smaller linewidth is attributed to an amorphous non-crystallized fraction of the embedded
MnO. This component does not show any anomalies around $T_N$. This is most probably due to the weakness of
interaction between the crystallized and amorphous fractions assuming that they occupy different voids in the
porous glass matrix.

On cooling the intensity of the dominating ESR signal from crystalline MnO increases and reaches a broad
maximum around 160-170 K. On further cooling towards to the N\'eel temperature the intensity decreases rapidly
(figure \ref{bulkvycor}c). In striking contrast to the bulk MnO, a strong weakly temperature dependent signal
remains below the onset of magnetic ordering and disappears only at low temperatures.

The ESR linewidth of this component on cooling in the paramagnetic region increases as in the bulk (figure
\ref{bulkvycor}d). However in contrast to the bulk, the linewidth does not exhibit a power-law divergence when
approaching $T_{N}$ from above. Instead a slight kink at about 150 K has been observed. It seems that the
antiferromagnetic critical fluctuations have been somehow suppressed in confinement. On cooling to temperatures
below the kink the ESR linewidth tends to continue to increase.

According to the neutron diffraction experiments \cite{MnO-PRL} the main difference between the bulk MnO and
MnO confined to a porous glass matrix is the existence of a large number of frustrated surface spins for the
latter case. The observation of the strong signal below the transition temperature, therefore is not surprising
for the magnetic nanoparticles and could be attributed to these "surface" disordered spins. Below the magnetic
transition ( $T_{N}$ = 122 K) the antiferromagnetic ordering occurs in the "core" part of nanoparticles, and
this part ceases to contribute to the ESR signal. The spins of the "surface" part of nanoparticles remain
disordered that results in the strong ESR signal even at low temperatures.

There is no theory describing the temperature dependence of the ESR linewidth for antiferromagnetic
nanoparticles and to the best of our knowledge, there exists only one experimental work focusing on this
subject. Sako et al. \cite{Sako} studied ESR spectra of the ultra fine nanoparticles of MnO, assembled in thin
layers embedded in the LiF matrix. However the ESR linewidth for ultra-fine nanoparticles was narrower by a
factor of 10 compared to our results and showed divergency at $T_N$ as in the bulk MnO. Accordingly the ESR
signal below the transition completely disappears. Our results are completely different to those of Sako et al.
This proves that the topology plays the extremely important role.

In spite of the old and rich history of antiferromagnetism in MnO the available experimental ESR data for the
bulk are rather controversial \cite{Drager,Battles,Dormann}. The universal formula for the temperature
dependence of the ESR linewidth $\Delta{H} \sim (T_{N} - T)^{- p}$, proposed for the short range ordered regime
\cite{Kawasaki} gives for the bulk MnO (figure \ref{bulkvycor}b) the effective parameters $T_N$ = 93(1) K and
\textit{p} = 1.08(2), while for confined MnO (figure \ref{bulkvycor}d) these parameters are: $T_{N}$ = 111(1) K
and \textit{p} = 0.56(5).

It is well known that the temperature dependence of the ESR linewidth in magnetic systems reflects the
development of the spin correlations\cite{Richard}. The differences in the critical exponents \textit{p} can
thus be also attributed to the differences in the development of the spin correlations imposed by a finite size
of nanoparticle. In this respect bulk MnO and MnO nanoparticles confined within porous glass show important
differences that need to be further theoretically investigated.

\subsection{MnO confined to channel-type matrices}

\begin{figure} [t]
\includegraphics* [width=\columnwidth] {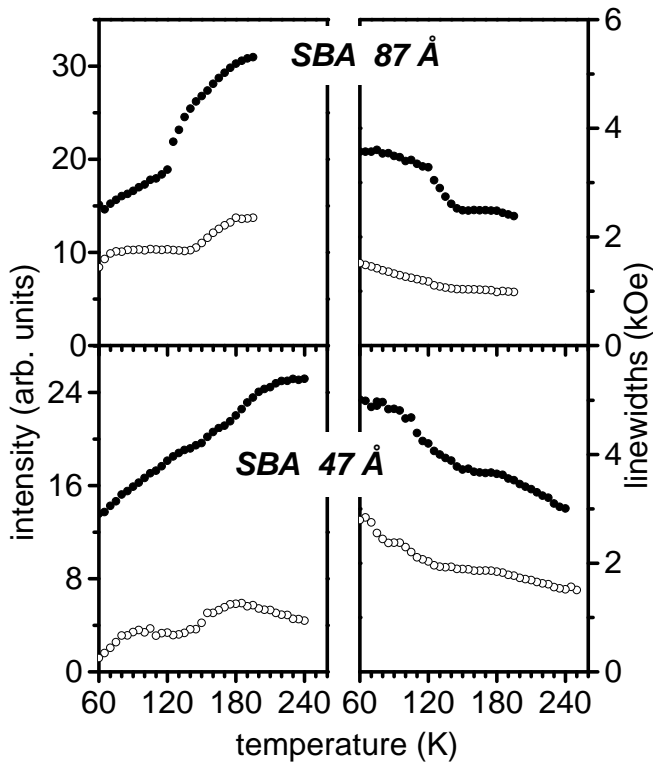}
\caption{Intensity (left column) and linewidth (right column) of the ESR signal from MnO confined to SBA
matrices with 87 {\AA} (upper row) and 47 {\AA} (lower row) channel diameter. Solid and open circles correspond
to the signals from amorphous and crystallized MnO, respectively. Errors (e.s.d.) do not exceed the symbol
size.} \label{sba}
\end{figure}

A different behavior was observed for MnO embedded into the channel-type matrices. In figure \ref{sba} the
intensities and linewidths of the signal components representative of amorphous and crystallized fractions of
confined MnO are shown for SBA with large channel diameters, which forms thin, ribbon-like structures.

Comparing the temperature dependencies of the ESR signal from crystallized MnO for channel type SBA and porous
glass (minor and main signals, respectively) it can be seen that they are similar. The ESR signal shows a
characteristic anomaly near the transition temperature that persists down to very low temperatures.

Surprisingly, the temperature dependencies of the ESR signal from the highly anisotropic nanoribbons with
reduced dimensionality within SBA is similar to the temperature dependency observed for the isotropic MnO
nanoparticles confined to a porous glass. This gives evidence that the magnetic order in nanoribbons is
substantially stabilized by the anisotropy and a low dimensionality has a small effect.

The ESR signal from the amorphous MnO in SBA gradually decreases with decreasing temperature that evidences the
development of the spin correlations in the "amorphous" part of the system. It is very likely that these
correlations are induced by the ordered magnetic system of the crystallized nanoparticles.

Indeed there is an anomaly associated with the onset of a long range magnetic ordering, observed in neutron
diffraction \cite{neutrons}. These traces of the "bulk behavior" in amorphous MnO are still seen for the matrix
with channel diameter of 87 {\AA}, while for the matrix with smaller diameter of 47 {\AA} they practically
disappear. Obviously these "traces" are associated with those spins in the amorphous fraction, which are close
to the crystallized nanoparticles with the strongly correlated moments.

\begin{figure} [t]
\includegraphics* [width=\columnwidth] {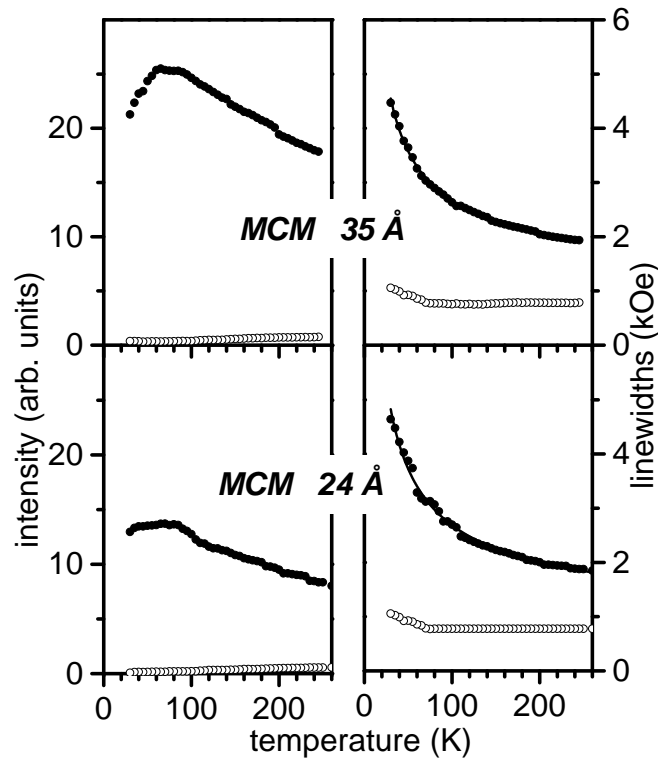}
\caption{Intensity (left column) and linewidth (right column) of the ESR signal from MnO confined to MCM
matrices with 35 {\AA} (upper row) and 24 {\AA} (lower row) channel diameter. Solid and open circles correspond
to the signals from amorphous and crystallized MnO, respectively. The weak signal from the crystallized MnO
within the matrix with 24 {\AA} channel diameter does not allow a reliable analysis. In this case the
refinement was carried out with a fixed linewidth, refined from the matrix with 35 {\AA} channel diameter.
Errors (e.s.d.) do not exceed the symbol size.} \label{mcm}
\end{figure}

In figure \ref{mcm} intensities and linewidths of the two constituents are shown for MCM with a smaller channel
diameter. In these matrices MnO forms nanowires with diameters smaller than 20 {\AA} and the lengths of about
180-200 {\AA}. Another distinguishing feature is that the ratio of crystalline phase to amorphous phase is an
order of magnitude smaller when compared with SBA (figure \ref{low-signal}).

As a result a small amount of the crystallized fraction in MCM with narrower channels is negligible in
comparison with an amorphous fraction. Therefore the temperature dependence of the ESR signal from the
amorphous fraction of MnO is typical for a system of disordered paramagnetic spins with a gradual increase of
the ESR signal up to the lowest temperatures (figure \ref{mcm}). Only at low temperatures the characteristic
"bump" indicates the appearance of spin correlations. The broad spectrum of relaxation rates, exchange
parameters and coordination numbers for Mn$^{2+}$ in this highly disordered system could explain high intensity
of the ESR signal and its linewidth divergence.

Because of low intensity any temperature peculiarities of the signal from crystallized nanoparticles cannot be
resolved within the limits of our experimental accuracy and we see only a monotonic decrease of the
corresponding component without any anomalies, which could manifest the onset of magnetic order at about 122 K,
clearly demonstrated by neutron diffraction \cite{neutrons}.

The linewidth divergence at $T = 0$ observed for the signal from crystallized, nanowire shape nanoparticles of
MnO means that the antiferromagnetic order does not occur until the lowest temperatures that should be
attributed to a low dimensionality of the magnetic system. Indeed, the observed temperature dependence of a
linewidth with a temperature independent "plateau", previous to the region of divergency (figure \ref{mcm}), is
similar to the behavior reported for low-dimensional systems, for example, for the one-dimensional
antiferromagnet Sr$_2$V$_3$O$_9$ \cite{Ivanshin}.

\section{Conclusion}

ESR experiments with antiferromagnet MnO confined within a porous vycor glass type and channel type matrices
have been interpreted on the basis of the structural data obtained by diffraction methods.

The fitting of the ESR signal from confined MnO requires two Lorentz components, while a signal from the bulk
is well fitted by a single Lorentzian. From comparison of the ESR data with the results of X-ray diffraction it
is shown that the signal component with the linewidth of about 1 kOe corresponds to crystallized MnO, while
another component is due to MnO in an amorphous state. Such analysis allows us to investigate the magnetic
behavior of the crystallized and amorphous parts of the embedded MnO separately.

We found that the signal from crystallized MnO dominates in porous glass, while in the case of channel type
matrices this signal is strongly suppressed on the account of a signal from an amorphous fraction of MnO. This
difference is attributed to the different pore topology for porous glass and channel type matrices. It leads to
a different "restricted geometry" for spins, in particularly, to different boundary surface dividing
magneto-ordered crystalline and disordered amorphous fractions.

The ESR signal associated with the crystalline MnO within a porous glass shows a behavior having many
similarities to the bulk behavior. However in contrast with the bulk the strong ESR signal due to disordered
"surface" spins is observed below the transition.

MnO crystallized within channel-type matrices possesses highly anisotropic, ribbon or wire-like structures
topologically very different from nanoparticles within a porous glass. Nevertheless the crystalline MnO phase
shares many characteristics with MnO in a porous glass, in particular a strong signal due to "surface" spins
below the magnetic transition still remains.

The mutual influence of amorphous and crystallized fractions is observed in the matrices with a larger channel
diameter. In this case the signal from the amorphous fraction shows progressive spin ordering with decreasing
temperature partially induced by ordered crystallized fraction.

With a decrease of channel diameter the fraction of the amorphous MnO increases while an amount of crystallized
MnO decreases. Therefore the ESR signal in matrices with a smaller channel diameter mainly originates from
amorphous MnO and its behavior is typical for the highly disordered magnetic system. The signal linewidth shows
a divergence at $T = 0$, while the ESR intensity demonstrates a broad maximum at low temperatures suggesting
the development of the spin correlation.

In conclusion, an amorphous MnO and crystalline MnO nanoparticles were identified and studied with the ESR
technique in nano-voids of different topology. The results underline the importance of the topological details
of the matrices and their influence on the magnetic properties. Our approach may pave the way to study the
topological aspects on the magnetism in confinement in a very systematic way.

\begin{acknowledgements}
The authors thank C. Alba-Simionesco, N. Brodie and G. Dosseh who prepared and characterized MCM and SBA
matrices. The authors are very grateful to I. Mirebeau for critical reading of manuscript and fruitful
discussion. The work was supported by the RFBR (Grant Nos. 02-02-16981 and 04-02-16550), the INTAS (Grant No.
2001-0826) and Russian-Slovenia Joint Research \& Technology Program. One of us, I.V.G., acknowledges the
financial support of Institute "Jozef Stefan", Ljubljana, Slovenia.
\end{acknowledgements}

\end{document}